\documentclass[
reprint,
superscriptaddress,
nofootinbib,
amsfonts,
amsmath,
amssymb,
aps,
prd,
floatfix
]{revtex4-2}

\usepackage{graphicx}
\usepackage{amsmath,amssymb,mathrsfs}
\usepackage{hyperref}
\usepackage{booktabs}
\usepackage{xcolor}

\begin{document}

\title{Characteristic evolution of conformal scattering:
\uppercase\expandafter{\romannumeral1}. Scalar Waves in Minkowski Spacetime}

\author{Zhen-Tao He}
\email{hezhentao22@mails.ucas.ac.cn}
\affiliation{School of Physical Sciences, University of Chinese Academy of Sciences, Beijing 100049, China}

\author{Yu Tian}
\email{ytian@ucas.ac.cn}
\thanks{Corresponding author}
\affiliation{School of Physical Sciences, University of Chinese Academy of Sciences, Beijing 100049, China}
\affiliation{\textit{Institute of Theoretical Physics, Chinese Academy of Sciences, Beijing 100190, China}}

\author{Hongbao Zhang}
\email{hongbaozhang@bnu.edu.cn}
\thanks{Corresponding author}
\affiliation{School of Physics and Astronomy, Beijing Normal University, Beijing 100875, China}
\affiliation{Key Laboratory of Multiscale Spin Physics, Ministry of Education, Beijing Normal University, Beijing 100875, China}

\date{\today}

\begin{abstract}
We study the conformal scattering of massless scalar waves in Minkowski spacetime.
The conformal scattering problem is formulated as a Goursat (characteristic initial-value) problem of the physical wave equation in compactified double-null coordinates, including the neighborhood of spatial infinity \(i^0\).
As null infinities \(\mathcal{I}^\pm\) lie on the domain boundary by construction, asymptotic radiation is directly accessible.
We consider three physical scenarios: free wave propagation, scattering off a P\"oschl--Teller (PT) potential, and the semi-linear \(|\phi|^{n-1}\phi\) wave equation.
For multipole numbers \(\ell =0,1\), an explicit stencil, averaging along the spatial direction, yields globally second-order convergent results.
For \(\ell \ge 2\), an implicit stencil averaging along the temporal direction is required for numerical stability.
Although the singular \(i^0\) reduces the convergence of the radiation data on \(\mathcal{I}^+\) to first order, Richardson extrapolation enhances the effective convergence rate to approximately \(1.5\).
For PT scattering, our method accurately computes scattering quantities, notably the phase shifts induced by the potential.
In the semi-linear case, our method captures the physical signatures of a self-defocusing Kerr nonlinearity, including self-phase modulation and spectral broadening.
The compactified double-null framework proves to be simple and efficient, suggesting a promising approach to the global evolution of conformal scattering.
\end{abstract}

\maketitle

\section{Introduction}
\label{sec:intro}

Scattering experiments serve as a powerful probe into the microscopic structures of a physical system and the underlying interaction mechanisms.
In scattering theory, scattering is formulated as a problem concerning the relation between asymptotically free states, i.e., ingoing states coming from the past and outgoing states receding toward the future.
For scattering problems in general relativity, Penrose, shortly after the foundational works by Bondi and collaborators on gravitational radiation \cite{Bondi:1962px,Sachs:1962wk}, developed an elegant conformal compactification method \cite{Penrose:1963}, with which the scattering problem takes a clean geometric form: the asymptotic characteristic initial data given at past null infinity $\mathcal{I}^-$, determine the induced radiation at future null infinity $\mathcal{I}^+$ \cite{Friedlander_1980,Hilditch:2019maz,Hilditch:2020dbh,Zhao:2021mez}.  
The conformal scattering has attracted considerable interest among mathematical relativists, who have been analyzing the corresponding scattering operator between Hilbert spaces on $\mathcal{I}^\pm$ \cite{HORMANDER1990270,doi:10.1142/S0219891604000123,Lindblad2023,kadar2025scatteringpolyhomogeneityasymptoticsquasilinear} (see \cite{Nicolas2024} for a brief review), especially asymptotic behaviors of solutions to the Goursat problem at the conformal boundary \cite{Valiente-Kroon:2002xys,ValienteKroon:2007azl,ValienteKroon:2008it,kroon2008regularityconditionsspatialinfinity,Gasperin:2021vnm,Minucci:2022hav,Compere:2023qoa,Ashtekar:2023wfn,taujanskas2023controlledregularityfuturenull,marajh2025controlledregularityfuturenull,Boschetti:2026ogm,Compere:2026jmk}. 

Numerical treatment is essential for tackling conformal scattering in more complicated settings.
To date, there are few mature numerical methods for handling global scattering evolution, primarily due to the singular nature of spatial infinity $i^0$ \cite{Friedrich_2018}, where generic solutions develop logarithmic singularities.
Several works \cite{Frauendiener:2013vda,FrauendienerHennig:2016,FrauendienerHennig:2017,Macedo:2018txl,Hennig:2020rns} developed fully spectral methods that obtain numerical solutions subject to certain regularity conditions on Friedrich's cylinder at $i^0$ \cite{Friedrich:1998}.
As for global solutions to the conformal scattering problem, Frauendiener and collaborators have numerically implemented Friedrich's Generalized Conformal Field Equations (GCFE) to study nonlinear perturbations of a black hole by gravitational waves \cite{FrauendienerStevens:2021,Frauendiener:2022bkj,FrauendienerGoodenbourStevens:2023,Camden:2025}.
In particular, they have recently achieved a fully nonlinear scattering simulation \cite{FrauendienerStevens:2025}, which represents a breakthrough in the scattering problem of general relativity.
While both null infinities $\mathcal{I}^\pm$ are covered in their GCFE method, the incoming scattering data are prescribed on a timelike boundary outside the physical spacetime and the spatial infinity $i^0$ is absent.\footnote{While we were finalizing this work, on arXiv appeared the work \cite{Barrer2026}, where Barrer and collaborators achieved fully global numerical evolutions of the linearised spin-2 field on conformally compactified Minkowski spacetime.} 

A different strategy has recently been proposed by Baptista and collaborators \cite{Baptista:2025kmo}, employing interpolation-connected future and past hyperboloidal foliations --- space-like hypersurfaces that asymptote to $\mathcal{I}^\pm$, respectively.
Although the neighborhood of $i^0$ is still absent in their hyperboloidal foliations, asymptotic data on $\mathcal{I}^\pm$ are directly accessible on the grid.
More recently, Demirbo\u{g}a and Zengino\u{g}lu \cite{DemirbogaZenginoglu:2026} have refined this hyperboloidal approach by introducing a central Penrose domain that bridges past and future hyperboloidal domains, which not only replaces interpolations between different domains with exact matching, but also includes the neighborhood of $i^0$.
However, their hyperboloidal methods are currently limited to scattering of scalar waves in Minkowski spacetime with spherical symmetry.
There are also hyperboloidal works \cite{zenginoğlu2021nullinfinitylayerwave,zenginoğlu2026penrosemelrosecomputingscattering,wess2026finiteelementshelmholtzscattering} by Zengino\u{g}lu and collaborators for scattering problems of the Helmholtz equation.

In this paper, we develop a compactified double-null framework to study the conformal scattering of scalar waves in Minkowski spacetime.
In our framework, null infinities \(\mathcal{I}^\pm\) are fixed on the domain boundary by construction, and no other unphysical domain is included, making asymptotic radiation directly accessible.
We find that this compactified double-null framework can handle non-spherically symmetric cases, and even black hole spacetimes (the conformal scattering of the Regge--Wheeler equation in Schwarzschild spacetime will be addressed in a forthcoming paper \cite{He2026}).
Even for the cubic case with a non-vanishing source at the conformal boundaries, our results exhibit global second-order convergence, without the deterioration of convergence reported in \cite{DemirbogaZenginoglu:2026}.
These facts suggest the compactified double-null framework is a promising approach to the global evolution of conformal scattering.

The paper is organized as follows.
In Sec.~\ref{sec:formalism}, we introduce the geometric setup for the conformal scattering problem in Minkowski spacetime, especially the physical wave equation in the compactified double-null coordinates $(U,V)$.
In Sec.~\ref{sec:numerics}, we present the numerical stencils for the characteristic integration, and discuss their stability and expected convergence rates in the interior and on $\mathcal{I}^+$.
In Sec.~\ref{sec:results}, we discuss numerical results for three physical scenarios: free scalar wave propagation in Sec.~\ref{sec:free}, scattering off the PT potential in Sec.~\ref{sec:pt_results}, and the semi-linear $|\phi|^{n-1}\phi$ wave equation in Sec.~\ref{sec:semilinear_results}.
We conclude in Sec.~\ref{sec:conclusion} with a summary of our findings and the outlook for future extensions.

\section{Geometric setup}
\label{sec:formalism}

\begin{figure}[htbp] 
\centering
\includegraphics[width=0.35\textwidth]{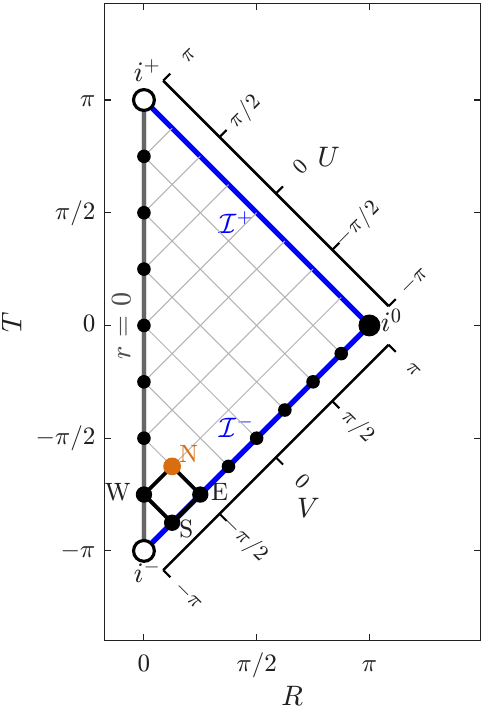}
\caption{Penrose diagram of Minkowski spacetime.
  Null infinities $\mathcal{I}^\pm$ are at $V=\pi$ and $U=-\pi$, respectively;
  $i^0$ and $i^\pm$ mark spatial and timelike infinities.
   Gray lines trace constant $U$ (slope $+1$) and constant $V$ (slope $-1$) characteristics.
  Black dots mark the grid points carrying prescribed data: the characteristic initial data on $\mathcal{I}^-$ (including $i^0$) and the regularity condition $\psi=0$ on $r=0$, while $i^\pm$ (open circles) require no data.
    A diamond-shaped cell with corners labeled E--S--W--N illustrates the numerical stencils for characteristic integration.
    }
\label{fig:minkowski_diagram}
\end{figure}

Starting from Minkowski spacetime in spherical coordinates,
\begin{equation}
  \mathrm{d}s^2 = -\mathrm{d}t^2 + \mathrm{d}r^2 + r^2 \mathrm{d}\Omega^2,
  \label{eq:minkowski}
\end{equation}
we introduce Penrose coordinates $(T,R)$ via
\begin{align}
      T& = \arctan(t+r) + \arctan(t-r),
      \\
  R &= \arctan(t+r) - \arctan(t-r),
\end{align}
and switch to compactified double-null coordinates
\begin{equation}
  U = T-R, \qquad V = T+R.
  \label{eq:UVdef}
\end{equation}
The physical domain $r\ge0$ corresponds to a triangle $0\le R<\pi\ \cap |T|+R<\pi$ in the Penrose coordinates.
The conformal factor reads 
\begin{equation}
  \Omega_{\text{P}} = \cos T + \cos R = 2\cos\frac{U}{2}\cos\frac{V}{2},
  \label{eq:OmegaP}
\end{equation}
which compactifies conformal infinities to the conformal boundary of the triangle ($\Omega_{\text{P}}=0$), see Fig.~\ref{fig:minkowski_diagram} for the locations of the conformal infinities in the Penrose diagram.

For the wave equation $\Box\phi = \mathcal{S}$ of a massless scalar field $\phi$ with source $\mathcal{S}$, we work with the physical rescaled field $\psi = r\phi$.
In compactified double-null coordinates $(U,V)$, the wave equation for $\psi$ reads
\begin{equation}
  \left(\partial_U\partial_V
    - \frac{\Delta_{\Omega}}{4\sin^2 R}\right)\psi
  = -\frac{r\mathcal{S}}{4\Omega_{\text{P}}^2},
  \label{eq:waveUV_full}
\end{equation}
where $R=(V-U)/2$ and $\Delta_{\Omega}$ is the Laplace--Beltrami operator on the unit sphere.  Expanding the field $\psi$ in terms of spherical harmonics $Y_{\ell m}$ as $\psi = \sum_{\ell m}\psi_{\ell m}(U,V)Y_{\ell m}$ and using $\Delta_{\Omega}Y_{\ell m} = -\ell(\ell+1)Y_{\ell m}$, the mode equation becomes
\begin{equation}
  \partial_U\partial_V
    \psi_{\ell m}
  = - \frac{\ell(\ell+1)}{4\sin^2 R}\psi_{\ell m} -\frac{r\mathcal{S}_{\ell m}}{4\Omega_{\text{P}}^2}.
  \label{eq:waveUV}
\end{equation}

The characteristic initial-boundary value problem under consideration is as follows: the characteristic initial data are prescribed on $\mathcal{I}^-$ (including $i^0$) and the scalar field $\phi$ is required to be regular at the origin $r=0$ (a diagonal line $V=U$ in the compactified double-null coordinates), namely, $\psi|_{r=0}=(r\phi)|_{r=0}=0$.
The setup of these prescribed data is shown in Fig.~\ref{fig:minkowski_diagram} as well.

In this paper, we adopt the oriented radiation fields convention defined in \cite{DemirbogaZenginoglu:2026}, i.e., $\mathcal{R}_- = \psi|_{\mathcal{I}^-},\, \mathcal{R}_+ = -\psi|_{\mathcal{I}^+}.$
The initial data on $\mathcal{I}^-$ are a modulated Gaussian pulse written in physical advanced time $v = \tan(V/2)$:
\begin{equation}  \label{eq:init}
\begin{aligned}
  \mathcal{R}_{-}(v) &= \psi(U=-\pi,\,V=2\arctan v) 
  \\
  &= A_0
  \exp\!\left(-\frac{v^2}{2\sigma_w^2}\right)\exp(-\mathrm{i}\omega_0 v),
 \end {aligned}
\end{equation}
with $A_0$ the amplitude,
$\sigma_w$ the width, and $\omega_0$ the modulation frequency.
The parameters are chosen so that the values of $\mathcal{R}_-$ at the grid points near $i^-$ and $i^0$ fall below machine precision, thereby rendering it numerically compactly supported.

We consider three cases in this paper:
\begin{itemize}
  \item {Free propagation:} $\mathcal{S} = 0.$

  \item {P\"oschl--Teller (PT) potential:}
    $\mathcal{S} =V_0\,\operatorname{sech}^2({r}/{\sigma})\,\phi.$
  \item {Semi-linear (defocusing):} $\mathcal{S} = \lambda |\phi|^{n-1}\phi, \quad \lambda>0.$
      
\end{itemize}
For the semi-linear case, we only consider the spherically symmetric mode $\ell=0$ here for simplicity, and the defocusing sign $\lambda>0$ ensures the energy is bounded below.

\section{Numerical Stencils for characteristic integration}
\label{sec:numerics}

Given the prescribed initial and boundary data described in Sec.~\ref{sec:formalism}, we can integrate the equation \eqref{eq:waveUV} throughout the spacetime and determine the outgoing radiation $\mathcal{R}_{+}$ on $\mathcal{I}^+$.
To do this, we discretize the $(U,V)$ domain with uniform spacing $\Delta = 2\pi/N$ and the four corners of each diamond cell $\mathcal{C}_{ij}:\;[U_{i-1},U_i]\times[V_{j-1},V_j]$ are denoted by $\psi_{\text{N}}=\psi(U_{i},V_{j}),\, \psi_{\text{E}}=\psi(U_{i-1},V_{j}),\, \psi_{\text{W}}=\psi(U_{i},V_{j-1})$, and $\psi_{\text{S}}=\psi(U_{i-1},V_{j-1})$, with $U_{i}=V_i=-\pi+i\Delta$ (see Fig.~\ref{fig:minkowski_diagram} for illustration).
In the following, we will introduce numerical stencils to solve the unknown value $\psi_{\text{N}}$ from the three known values $\psi_{\text{E}}$, $\psi_{\text{W}}$, and $\psi_{\text{S}}$.

Recasting the partial differential equation (PDE) \eqref{eq:waveUV} as $\partial_U\partial_V\psi = -S(U,V;\psi)$ \footnote{From now on, we drop the subscript $\ell m$ for simplicity.} and integrating it over the diamond cell $\mathcal{C}_{ij}$ give
\begin{equation}
  \psi_{\text{N}} - \psi_{\text{E}} - \psi_{\text{W}} + \psi_{\text{S}}
  = -\iint_{\mathcal{C}} S(U,V;\psi)\;\mathrm{d}U\,\mathrm{d}V.
  \label{eq:integral}
\end{equation}
The integral at right-hand side is approximated by the midpoint rule
\begin{equation}
    \iint_{\mathcal{C}_{ij}} S(U,V;\psi)\;\mathrm{d}U\,\mathrm{d}V
    = 
    S(U_{\rm c},V_{\rm c};\psi_{\rm c}) \Delta^2 + I_\epsilon
    ,
\end{equation}
where $U_{\rm c}= (U_i+U_{i-1})/2,\, V_{\rm c}=(V_j+V_{j-1})/2$ and $\psi_{\rm c}=\psi(U_{\rm c},V_{\rm c})$. 
The choice of center approximation for $\psi_{\text{c}}$ affects numerical stability and convergence.
Specifically, we use the following explicit or implicit stencil based on the multipole number $\ell$.
\begin{itemize}
    \item \textbf{Explicit Stencil ($\ell=0,1$)}:
The center value $\psi_{\rm c}\simeq(\psi_{\text{E}}+\psi_{\text{W}})/2$ is approximated by the average of the two known values along the spatial direction.
\item \textbf{Implicit Stencil ($\ell\ge2$).}
The center value $\psi_{\rm c}\simeq(\psi_{\text{N}}+\psi_{\text{S}})/2$ is approximated by the average of the two corners along the temporal direction.
\end{itemize}
The convergence order of the midpoint rule depends on the regularity of $S(U,V;\psi)$.
If $S(U,V;\psi)$ is a $C^n$ function with $n\ge2$ in $\mathcal{C}_{ij}$, then the truncation error $I_\epsilon\sim \mathcal{O}(\Delta^4)$, yielding a second-order method.
Otherwise, the singularity of $S(U,V;\psi)$ deteriorates the order of the midpoint rule. 
As we will see in the following numerical experiments in Sec.~\ref{sec:results}, the convergence of $\mathcal{R}_+$ at future null infinity $\mathcal{I}^+$ is reduced to first-order by the singular $i^0$ when $\ell\ge2$.

Terms on the right-hand side of Eq.~\eqref{eq:waveUV} have different singular behaviors in the Penrose triangle.
Despite being avoided by the midpoint rule, these singular points could lead to not only weaker convergence of numerical methods, but also numerical instability.
Specifically, the effective centrifugal potential $V_\ell(R)=\ell(\ell+1)/(4\sin^2 R)$ diverges when $\sin R=0$, that is, at $R=0$ (including $r=0$ and timelike infinities $i^\pm$) and $R=\pi$ (i.e., spatial infinity $i^0$).
For the PT potential, the effective potential $V_{\rm PT}=V_0\,\operatorname{sech}^2({r}/{\sigma})/(4\Omega_{\rm P}^2)$ diverges at the timelike infinities $i^\pm$.
The semi-linear effective source term $-{\lambda\,\Omega_\text{P}^{n-3}} |\psi|^{n-1}\psi/{4\sin^{n-1}R}$ seems intractable, but numerically, it appears regular, causing no instability or deterioration of convergence.

For the linear wave equation with an effective potential $\partial_U\partial_V\psi =-V_{\rm eff}\psi$, the explicit stencil corresponds exactly to the Gundlach–Price–Pullin scheme \cite{Gundlach:1993}:
\begin{equation}
   \psi_{\text{N}}=\left(1- \frac{c}{2}\right)(\psi_{\text{E}} + \psi_{\text{W}})  - \psi_{\text{S}}
   ,
  \label{eq:explicit}
\end{equation}
with $c \equiv \Delta^2 V_{\text{eff}}(R_{\text{c}})$ and $R_{\text{c}} = (V_j+V_{j-1}-U_i-U_{i-1})/4$ evaluated at the diamond center.
A local Courant-Friedrichs-Lewy (CFL) analysis shows that the coefficient of $(\psi_{\text{E}}+\psi_{\text{W}})$ serves as an amplification factor.
Thus, a sufficient condition for the stability of the explicit stencil is $|1-c/2|\le1$, which requires $0 \le c \le 4$.
On the other hand, solving the implicit stencil of the linear wave equation for $\psi_{\text{N}}$ yields an explicit formula
\begin{equation}
  {\psi_{\text{N}} = \frac{\psi_{\text{E}} + \psi_{\text{W}}}{1 + c/2}
    - \psi_{\text{S}}}.
  \label{eq:implicit}
\end{equation}
The amplification factor $1/(1+c/2)\in(0,1]$ for all $c\ge0$, guaranteeing unconditional stability.
For the free case $V_{\rm eff}=V_\ell$, the maximum value $c_{\max}\approx\ell(\ell+1)$ occurs at the first off-diagonal cells $\mathcal{C}_{i,i+1},i=1,\dots,N$ near $R=0$ and the cell $\mathcal{C}_{1N}$ including $i^0$ (where $R=\pi$). 
In these regions, $\sin R_{\rm c}\simeq\Delta/2$, which yields $V_{\ell}\approx \ell(\ell+1)/\Delta^2$ and hence $c\approx\ell(\ell+1)$.
Consequently, the explicit stencil is numerically stable for $\ell=0,1$ (where $0<c \le c_{\max} =2 < 4$).
\footnote{We find empirically that the violation of the local CFL condition for $\ell>1$ usually, but not always, leads to instability.
We checked up to $\ell=6$ and found that $\ell=3$ is the only stable case.}
When the PT potential is included ($V_{\rm eff}=V_\ell+V_{\rm PT}$), timelike infinities become more singular, where the explicit stencil becomes unstable. 
One can just cut off several cells near $i^\pm$ where $c>4$ or use the implicit stencil there instead.
The reason for preserving the explicit stencil is that for $\ell=1$, we find empirically that the explicit stencil achieves global second-order convergence, but the radiation $\mathcal{R}_+$ obtained by the implicit stencil exhibits only first-order convergence, indicating $\psi_{\rm c}\simeq(\psi_{\text{E}}+\psi_{\text{W}})/2$ is a better approximation here.

For the semi-linear case, a rigorous local CFL analysis is complicated by the nonlinear source terms; however, we find empirically that the explicit stencil remains numerically stable.

\section{Results}
\label{sec:results}
In this section, we show numerical results obtained by the characteristic integration stencils, report their convergence, and check them against exact or known results. 
First, the results of free propagation are shown in Sec.~\ref{sec:free}, checked by a parity relation $\mathcal{R}_{-}(s)=(-1)^\ell\mathcal{R}_+(s)$.
We also introduce self-convergence as a numerical diagnostic and Richardson extrapolation in this subsection.
Second, in Sec.~\ref{sec:pt_results}, scattering of the PT potential is checked by phase shifts and the elasticity of the scattering.
Third, the results of the semi-linear $|\phi|^{n-1}\phi$ wave equation are shown in Sec.~\ref{sec:semilinear_results}, examined by energy conservation deviation, where we also observe spectral broadening in the spectrum of $\mathcal{R}_+(u)$, namely a Kerr effect.\footnote{
Recently, \cite{cardoso2026nonlineardynamicsgeneralrelativity} studied nonlinear scattering of scalar waves in flat space in a different context, and also reported a gravitational Kerr effect.}

\subsection{Free propagation}
\label{sec:free}

\begin{figure}[htbp] 
\centering
\includegraphics[width=0.24\textwidth]{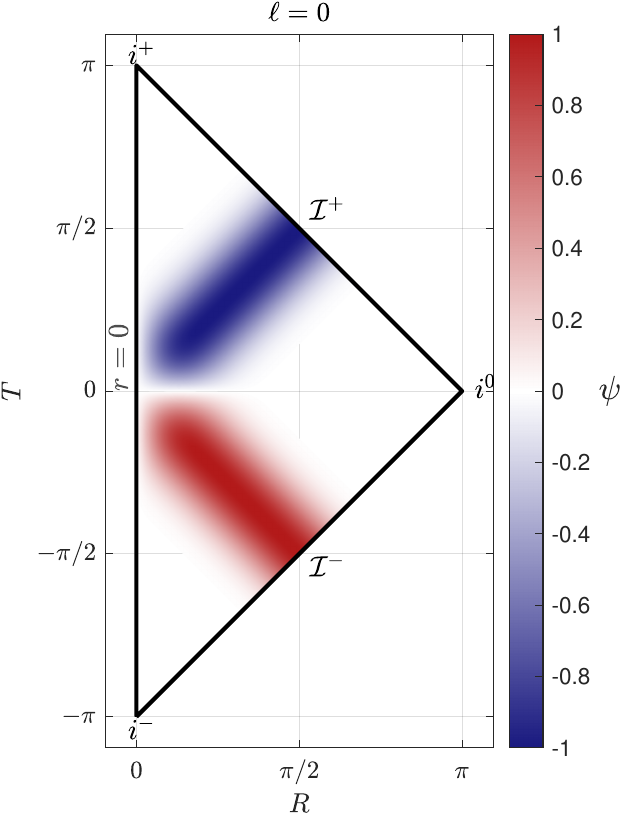}\hfill
\includegraphics[width=0.24\textwidth]{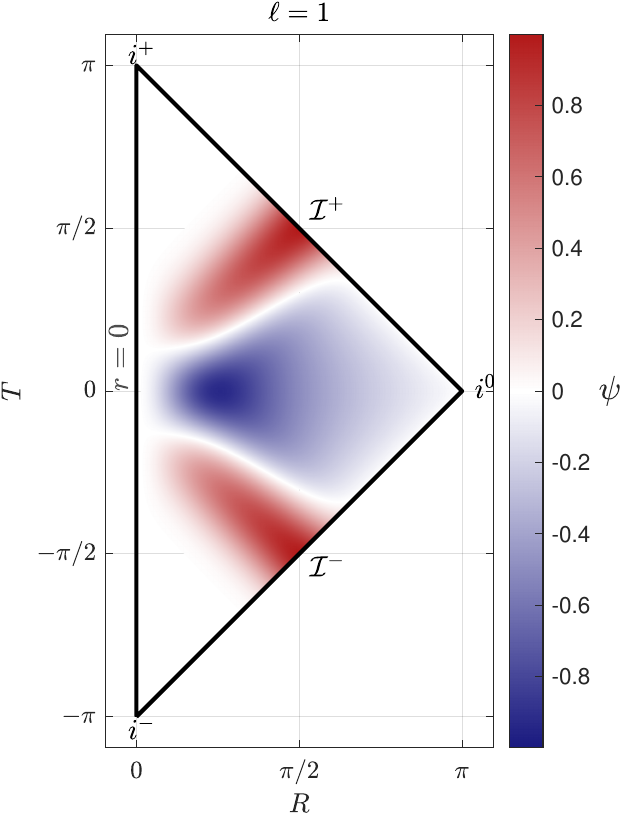}\hfill
\includegraphics[width=0.24\textwidth]{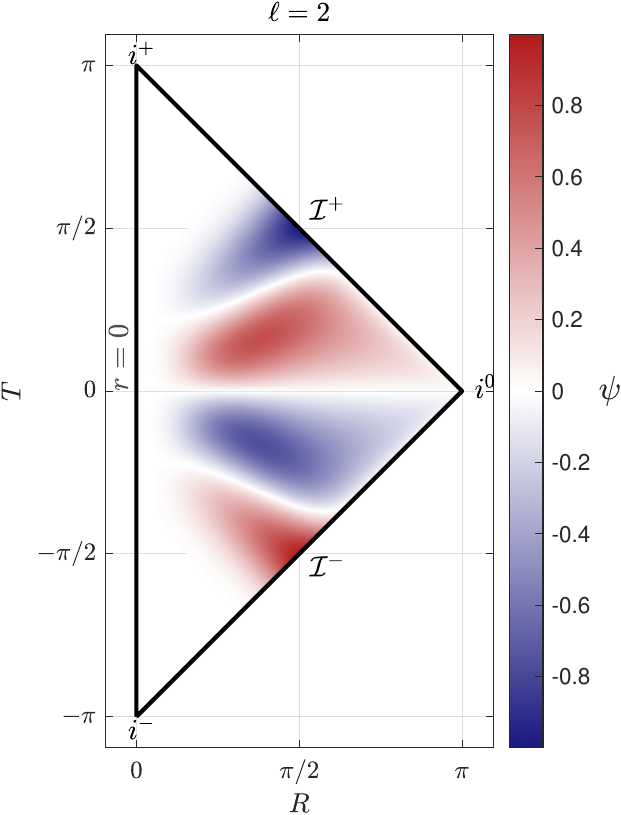}\hfill
\includegraphics[width=0.24\textwidth]{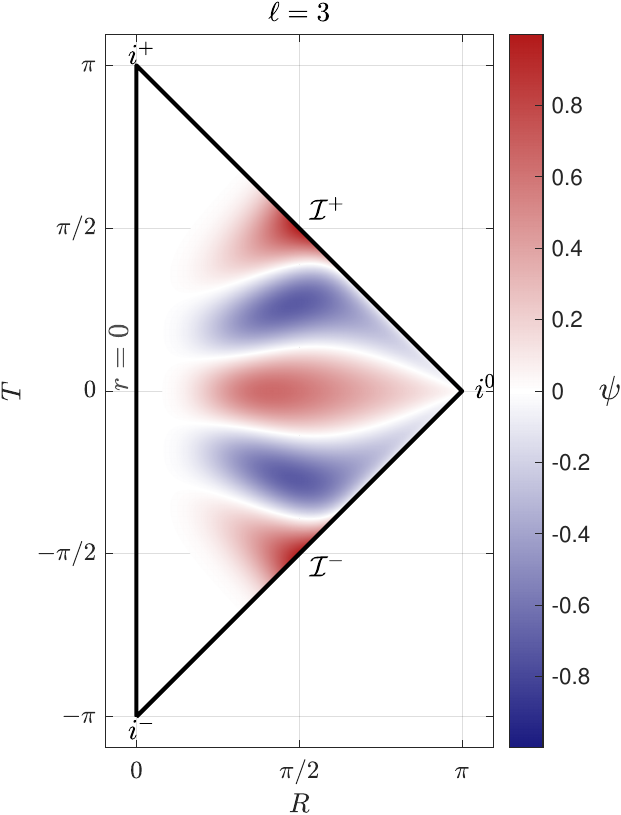}
\caption{Penrose diagrams of the free propagation for $\ell=0,1,2,3$
  ($A_0=1$, $\sigma_w=0.2,\,\omega_0=0$, $N=4096$).}
\label{fig:free_l03}
\end{figure}

Fig.~\ref{fig:free_l03} shows the free propagation of scalar waves on
Minkowski spacetime.
For $\ell=0$, $V_{\ell}\equiv0$ and both the explicit stencil \eqref{eq:explicit} and the implicit stencil \eqref{eq:implicit} reduce to the exact discrete integral of the initial data.
Therefore, the error against the exact analytic free wave solution is dominated by round-off error.

When there is no analytic solution for reference, we compute self-convergence rates by comparing solutions with successive resolution doublings ($N,2N,4N$):
\begin{equation}
    Q_{2/\infty}(\psi;N) = \log_2
    \left(
    \frac{\|e_{N,2N}\|_{2/\infty}}{\|e_{2N,4N}\|_{2/\infty}}
    \right),
\end{equation}
where $\|e_{N_1,N_2}\|_{2/\infty}$ denote the discrete $L_2$ and $L_\infty$ norms of $\psi_{N_1}-\psi_{N_2}$ over the grid points shared by all resolutions, respectively.  
In this paper, we are interested in convergence of data in the global computation domain (excluding the prescribed data) and $\mathcal{I}^+$.

For $\ell=1$, both the $L_2$ and $L_\infty$ errors exhibit second-order convergence in the global computation domain.
For $\ell\ge2$, the global convergence order of the $L_2$ and $L_\infty$ errors is deteriorated by the errors near $\mathcal{I}^+$.
For example, Table~\ref{tab:convl2} shows the global convergence order $Q_{2/\infty}(\psi;N=1024)$ for $\ell=2$, where a recovery of second-order convergence is observed by excluding $k$ layers of data near $V_k=\pi-k\cdot 2\pi/N$.

\begin{table}[htbp] 
\centering
\caption{Convergence order $Q_{2/\infty}(\psi;N=1024)$ for $\ell=2$ when excluding $k$ layers $V_k=\pi-k\cdot 2\pi/N$ from $\mathcal{I}^+$ ($V=\pi$).}
\begin{tabular}{lcc}
\toprule
$k$ layers & $Q_2$  & $Q_\infty$  \\
\midrule
$k=0$ (global) & 1.52 & 1.00 \\
$k=1$         & 2.00 & 1.78 \\
$k=2$         & 2.04 & 1.81 \\
$k=3$         & 2.04 & 1.94 \\
\bottomrule
\end{tabular}
\label{tab:convl2}
\end{table}

\begin{figure*}[htbp] 
\centering
\includegraphics[width=0.85\textwidth]{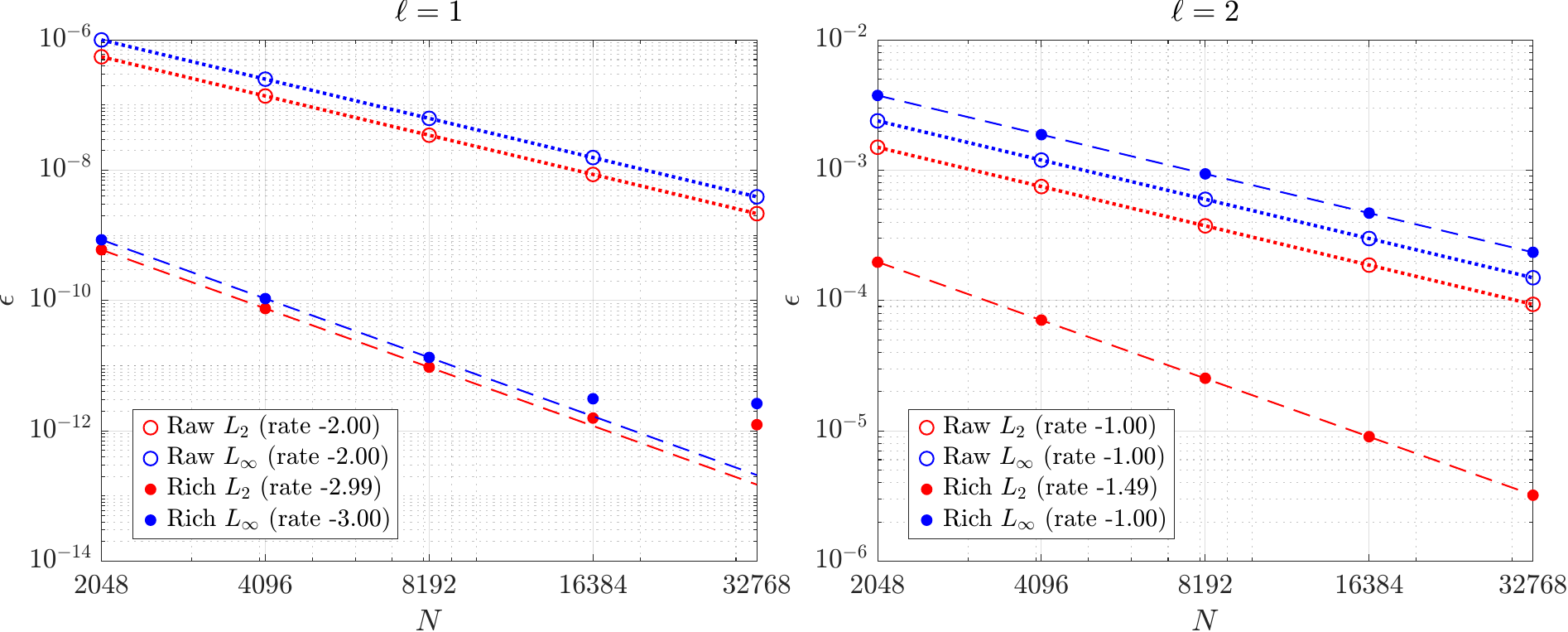}
\caption{The $L_{2/\infty}$ norm of the error $\epsilon = |\mathcal{R}_{-}(s)-(-1)^\ell\mathcal{R}_+(s)|$ of raw $\mathcal{R}_{+,N}$ and Richardson-extrapolated $\mathcal{R}_{+,N/2}^{[p]}$ for $\ell=1$ ($p=2$) and $\ell=2$ ($p=1$).
These are log-log scale plots, where fitted dashed lines exhibit algebraic convergence $\sim N^{\rm rate}$.
Note that $\mathcal{R}_{+,N/2}^{[p]}$ is obtained from the raw results $\mathcal{R}_{+,N}$ and $\mathcal{R}_{+,N/2}$.
For $\ell=1$, the $L_{2}$ and $L_{\infty}$ norms of $\epsilon$ of $\mathcal{R}_{+,N/2}^{[p]}$ are much less than those of $\mathcal{R}_{+,N}$.
For $\ell=2$, however, the $L_{2}$ norm of $\epsilon$ of $\mathcal{R}_{+,N/2}^{[p]}$ is a bit less than that of $\mathcal{R}_{+,N}$, and the $L_{\infty}$ norm is slightly increased after extrapolation.
Also note that for $\ell=1$, the dashed lines are fitted by data of $N=2048,4096$ and $8192$, and data of $N=16384$ and $32768$ are dominated by accumulated round-off errors, thereby deviating from the power law.
}
\label{fig:richardson}
\end{figure*}

Although the convergence order of the radiation $\mathcal{R}_+$ is poor, the leading order error $\mathcal{O}(\Delta^p)$ can be suppressed by Richardson extrapolation:

\begin{equation}
  \mathcal{R}_{+,N}^{[p]}
  = \frac{2^p \mathcal{R}_{+,2N} - \mathcal{R}_{+,N}}{2^p-1}.
\end{equation}
As shown in Fig.~\ref{fig:richardson}, the $L_2$ and $L_\infty$ norms of the error defined by the parity relation $\epsilon = |\mathcal{R}_{-}(s)-(-1)^\ell\mathcal{R}_+(s)|$ converge faster after Richardson extrapolation.
For $\ell=1$, the leading order of the raw error is $\sim\mathcal{O}(\Delta^2)$, and the error after extrapolation is dominated by the sub-leading order $\sim\mathcal{O}(\Delta^3)$, suggesting a $C^2$ regularity of $V_{\ell=1}\psi$ near $V=\pi$.
For $\ell=2$, the $L_2$ rate improves from $1.0$ to ${\sim}1.5$ after extrapolation, while the $L_\infty$ rate remains ${\sim}1.0$.
The cases for $\ell>2$ are nearly the same as $\ell=2$.  

\subsection{Scattering off the PT potential}
\label{sec:pt_results}

\begin{figure}[htbp] 
\centering
\includegraphics[width=0.24\textwidth]{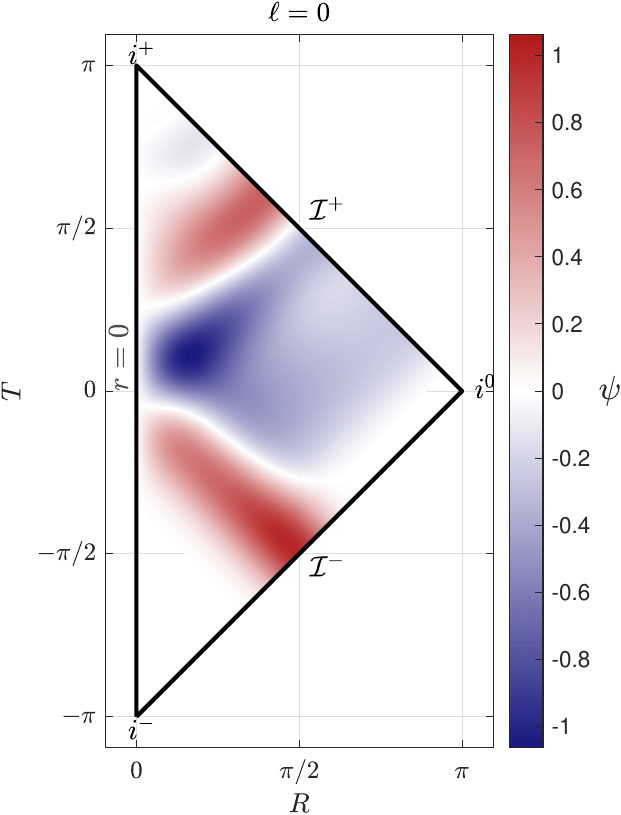}\hfill
\includegraphics[width=0.24\textwidth]{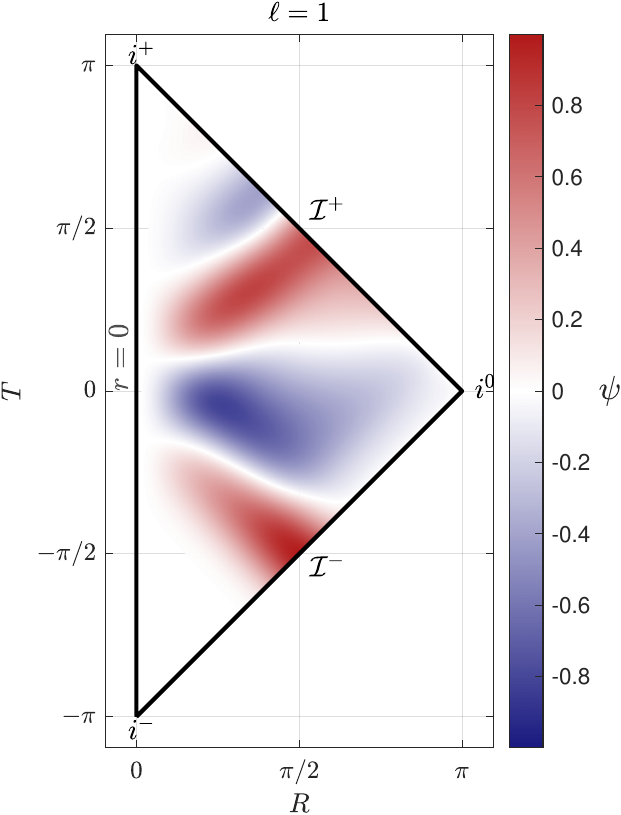}\hfill
\includegraphics[width=0.24\textwidth]{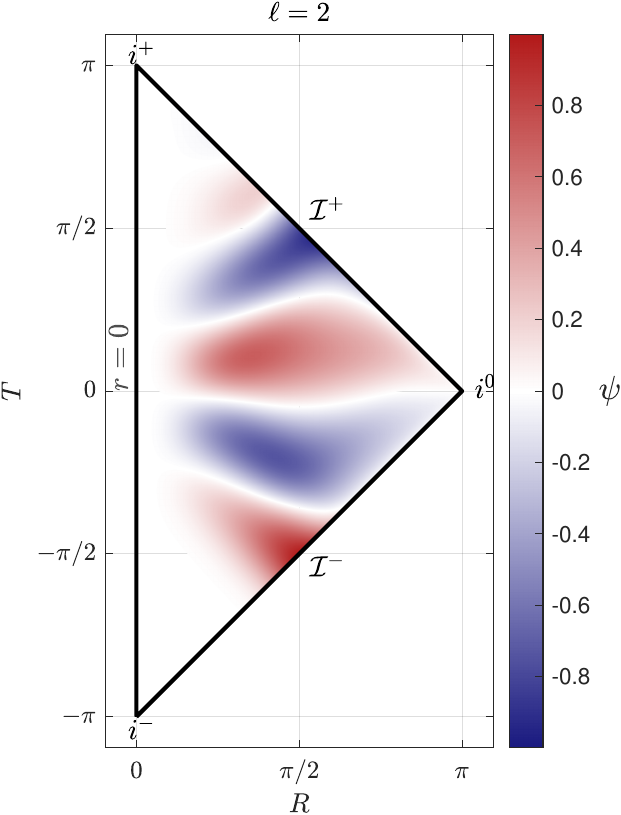}\hfill
\includegraphics[width=0.24\textwidth]{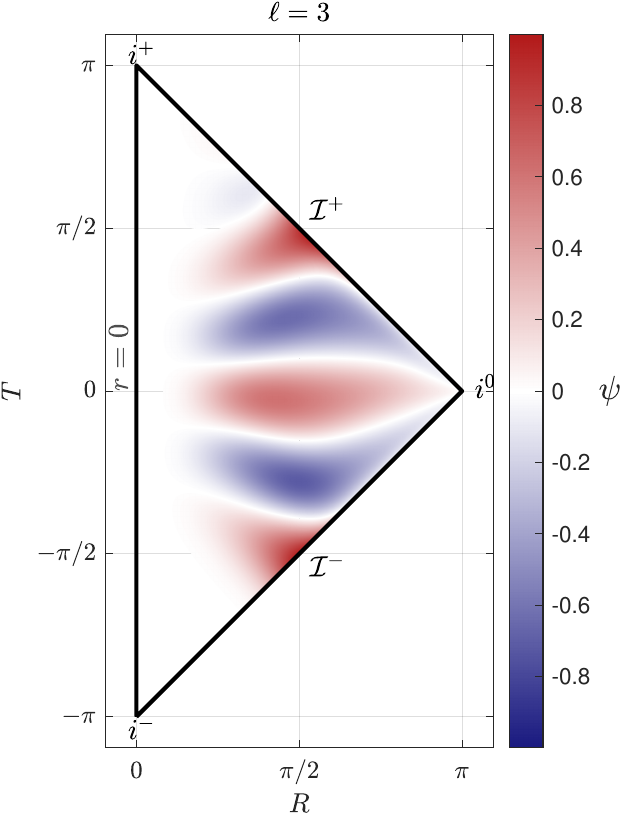}
\caption{Penrose diagrams of scattering off the PT potential for $\ell=0,1,2,3$ ($A_0=1$, $\sigma_w=0.2,\,\omega_0=0$, $V_0=5$, $\sigma=1$, $N=4096$).}
\label{fig:pt_penrose}
\end{figure}

\begin{figure}[htbp] 
\centering
\includegraphics[width=0.45\textwidth]{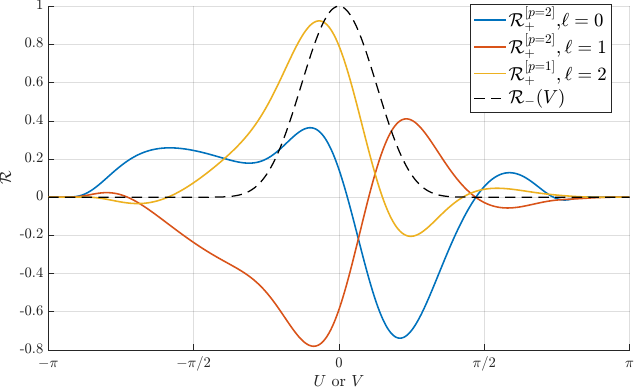}
\caption{Richardson-extrapolated radiation $\mathcal{R}_{+,N}^{[p]}$ for scattering off the PT potential when $\ell=0,1,2$ (other parameters are the same as Fig.~\ref{fig:pt_penrose}).}
\label{fig:pt_rad}
\end{figure}

\begin{figure*}[htbp] 
\centering
\includegraphics[width=\textwidth]{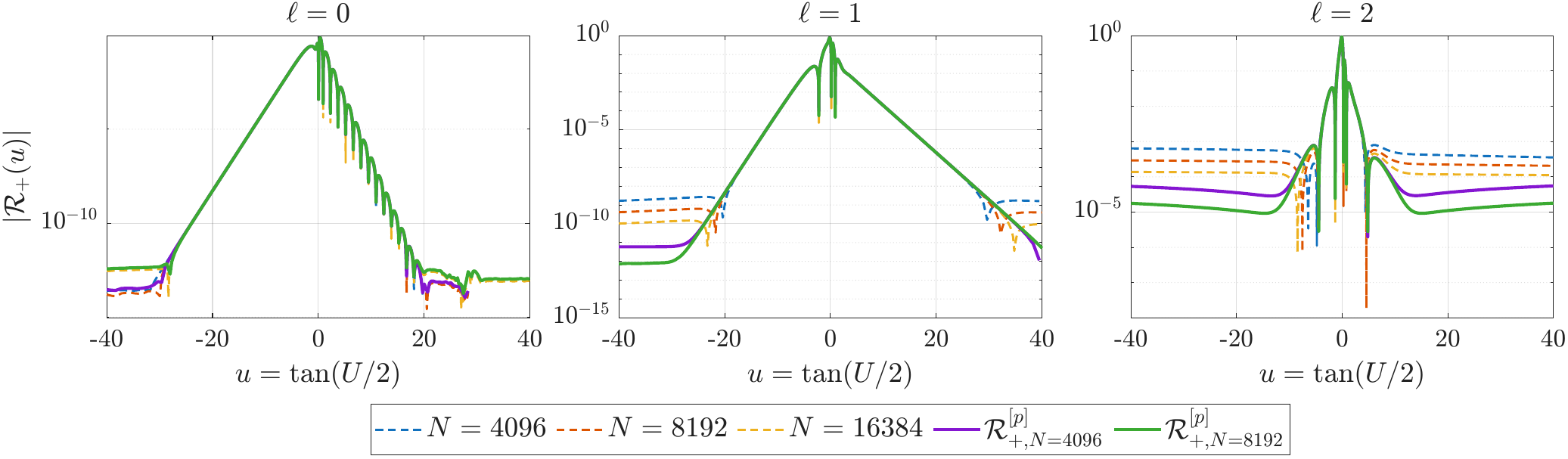}
\caption{Semilog plots of raw $\mathcal{R}_{+,N}$ and Richardson-extrapolated $\mathcal{R}_{+,N}^{[p]}$ for scattering off the PT potential with different resolutions $N$ when $\ell=0,1,2$ (other parameters are the same as Fig.~\ref{fig:pt_penrose}).
The extrapolation order is $p=2$ for $\ell=0,1$ and $p=1$ for $\ell=2$.
Left panel ($\ell=0$): only quasi-normal ringing.
Middle panel $(\ell=1)$: exponential tail decay after transient quasi-normal ringing.
Right panel $(\ell=2)$: also suggesting exponential tail decay after transient quasi-normal ringing, but higher resolution is needed to capture the tail behavior precisely due to the slow convergence.}
\label{fig:pt_rd}
\end{figure*}

The conformal scattering of the PT potential is shown in Fig.~\ref{fig:pt_penrose}, and the radiation $\mathcal{R}_+$ is shown in Fig.~\ref{fig:pt_rad} and Fig.~\ref{fig:pt_rd}.
For $\ell=0,1$, the explicit stencil achieves global $O(\Delta^2)$ convergence. 
For $\ell\ge2$, however, the implicit stencil gives interior $O(\Delta^2)$ convergence and $O(\Delta)$ convergence on $\mathcal{I}^+$, the same as the free case.
Table~\ref{tab:pt_conv} lists the self-convergence rates of the radiation $\mathcal{R}_+$ for raw and Richardson-extrapolated data when $\ell=0,1,2$.
For $\ell=0,1$, both the $Q_2$ and $Q_\infty$ are increased after the Richardson extrapolation with $p=2$, where a superconvergence $(Q_2\to\sim8.0,\,Q_\infty\to\sim9.5)$ for $\ell=0$ is observed. 
For $\ell=2$, the Richardson extrapolation with $p=1$ only improves the $L_2$ rate $Q_2$ from $1.0$ to ${\sim}1.5$, which is the same for $\ell>2$.

\begin{table*}[htbp] 
\centering
\caption{Self-convergence rates of raw $\mathcal{R}_+$ and Richardson-extrapolated $\mathcal{R}^{[p]}_{+,N}$ for PT scattering
  ($\ell=0,1,2$, $A_0=1$, $\sigma_w=0.2$, $\omega_0=0$, $V_0=5$, $\sigma=1$).
  Rates are computed as $Q=\log_2(\|e_{N,2N}\|/\|e_{2N,4N}\|)$.}
\label{tab:pt_conv}
\begin{tabular}{lcccccccccccc}
\toprule
 & \multicolumn{6}{c}{$Q_{2/\infty}(\mathcal{R}_+;N=2048)$} & \multicolumn{6}{c}{$Q_{2/\infty}(\mathcal{R}^{[p]}_{+,N};N=1024)$} \\
\cmidrule(lr){2-7}\cmidrule(lr){8-13}
$\ell$ & $Q_2$ & $\|e_{N,2N}\|_2$ & $\|e_{2N,4N}\|_2$ & $Q_\infty$ & $\|e_{N,2N}\|_\infty$ & $\|e_{2N,4N}\|_\infty$
       & $Q_2$ & $\|e_{N,2N}\|_2$ & $\|e_{2N,4N}\|_2$ & $Q_\infty$ & $\|e_{N,2N}\|_\infty$ & $\|e_{2N,4N}\|_\infty$ \\
\midrule
0 & \textbf{2.00} & $8.7\!\times\!10^{-7}$ & $2.2\!\times\!10^{-7}$ & \textbf{2.00} & $1.8\!\times\!10^{-6}$ & $4.6\!\times\!10^{-7}$
  & \textbf{7.98} & $4.9\!\times\!10^{-9}$ & $1.9\!\times\!10^{-11}$ & \textbf{9.50} & $1.1\!\times\!10^{-7}$ & $1.5\!\times\!10^{-10}$ \\
1 & \textbf{2.00} & $5.6\!\times\!10^{-7}$ & $1.4\!\times\!10^{-7}$ & \textbf{2.00} & $1.3\!\times\!10^{-6}$ & $3.2\!\times\!10^{-7}$
  & \textbf{2.98} & $3.4\!\times\!10^{-10}$ & $4.3\!\times\!10^{-11}$ & \textbf{3.11} & $4.2\!\times\!10^{-10}$ & $4.9\!\times\!10^{-11}$ \\
2 & \textbf{1.00} & $5.0\!\times\!10^{-4}$ & $2.5\!\times\!10^{-4}$ & \textbf{1.00} & $8.6\!\times\!10^{-4}$ & $4.3\!\times\!10^{-4}$
  & \textbf{1.49} & $9.8\!\times\!10^{-5}$ & $3.5\!\times\!10^{-5}$ & \textbf{1.00} & $2.0\!\times\!10^{-3}$ & $9.9\!\times\!10^{-4}$ \\
\bottomrule
\end{tabular}
\end{table*}

For the linear wave equation with a real static potential, the scattering is purely elastic, and the scattering information is encoded in the frequency-dependent phase shift $\delta_\ell(\omega)$.
\footnote{
Quasi-normal modes (QNMs) or late-time tails (e.g., shown in Fig.~\ref{fig:pt_rd}) are also important for checking the validity of numerical methods.
However, they are insufficient to validate the accuracy of numerical experiments for conformal scattering, because QNMs and tails, which can be excited by many types of initial data, originate from analytic properties of the retarded Green's function (poles and branch cuts), not from the scattering operator.
}
We define the transfer function $H_\ell(\omega)=\tilde{\mathcal{R}}_+(\omega)/\tilde{\mathcal{R}}_-(\omega)$ as the quotient of Fourier transforms of $\mathcal{R}_+(u)$ and $\mathcal{R}_-(v)$.
The phase shift is defined as $\delta_\ell(\omega)=\arg H_\ell(\omega)$ here, and a transfer function $|H_\ell(\omega)|=1$ with unit modulus represents the elastic scattering.

We extract $H(\omega)$ via Fast Fourier Transform (FFT)\footnote{The discrete Fourier transform implemented here is defined as $Y_k=\sum_{j=1}^{n}X_j\exp[-2\pi\mathrm{i}(j-1)(k-1)/n]$ for two vectors $X$ and $Y$ of length $n$.} of $\mathcal{R}_+(u)$ and $\mathcal{R}_-(v)$.  
From $\mathcal{R}_+(U)$ on the uniform $U$-grid, we map to the physical retarded time $u = \tan(U/2)$.
Near the corners $U\to\pm\pi$, the local $u$-grid spacing $\delta u$ grows rapidly, making the non-uniform sampling too sparse for reliable interpolation.
We retain the interval where $\delta u < (\delta u)_{\max}$, with $(\delta u)_{\max}$ chosen by hand so that the retained grid density is sufficient for the shape-preserving pchip (piecewise cubic Hermite interpolating polynomial \cite{Fritsch:1980}) interpolation onto a uniform $u$-grid.
The resampled signal is then Fourier-transformed via FFT without windowing or zero-padding.
The incoming spectrum $\tilde{\mathcal{R}}_{-}(\omega)$ is obtained by applying the identical procedure to the analytic initial data on $\mathcal{I}^-$.  
\begin{equation}
  \tilde{\mathcal{R}}_{-}(\omega) = A_0\sigma_w
  \exp\!\left[-\frac{1}{2}\sigma_w^2(\omega-\omega_0)^2\right].
  \label{eq:fftR-}
\end{equation}

To check the phase shifts $\delta_\ell^{\rm PDE}(\omega)$ obtained by characteristic integration of the PDE \eqref{eq:waveUV} at the discrete FFT frequencies, we solve the ordinary differential equation (ODE) for the radial wave function $\psi_\omega(r)=\mathrm{e}^{-\mathrm{i}\omega t}\psi(t,r)$
\begin{equation}
  \left[\partial_r^2+\omega^2-\frac{\ell(\ell+1)}{r^2} -V_0\operatorname{sech}^2\left(\frac{r}{\sigma}\right)
  \right]
  \psi_\omega=0,
  \label{eq:freq_ode}
\end{equation}
where $\psi_\omega$ satisfies the boundary conditions $\psi_\omega(0)=0$ and $\psi_\omega(r\to\infty)\to A_{\rm out} \mathrm{e}^{\mathrm{i}\omega r} + A_{\rm in} \mathrm{e}^{-\mathrm{i}\omega r}$.
Then the $S$-matrix element $S_\ell(\omega)=A_{\rm out}/A_{\rm in}$ gives the transfer function $H_\ell(\omega)=-\overline{S_\ell(\omega)}$. 

For $\ell=0$, an exact analytic expression for $S_0(\omega)$ follows from the known hypergeometric solution of the PT potential~\cite{Flugge71}
\begin{equation}
  S_0(\omega)=\frac{\Gamma(\mathrm{i}\omega/\alpha)}{\Gamma(-\mathrm{i}\omega/\alpha)}
  \frac{\Gamma(B)\,\Gamma(C-A)}{\Gamma(A)\,\Gamma(C-B)}\,
  2^{-2\mathrm{i}\omega/\alpha},
  \label{eq:Sanalytic}
\end{equation}
with $\alpha=1/\sigma$, $\lambda(\lambda-1)=-V_0\sigma^2$, and $A,B=(\lambda+1)/2\pm i\omega/2\alpha$, $C=3/2$.
For $\ell\ge1$, no closed-form analytic expression is known; therefore, the phase shift is obtained numerically via the Chebyshev spectral method described in Appendix~\ref{app:cheb}.

\begin{figure}[htbp] 
\centering
\includegraphics[width=0.45\textwidth]{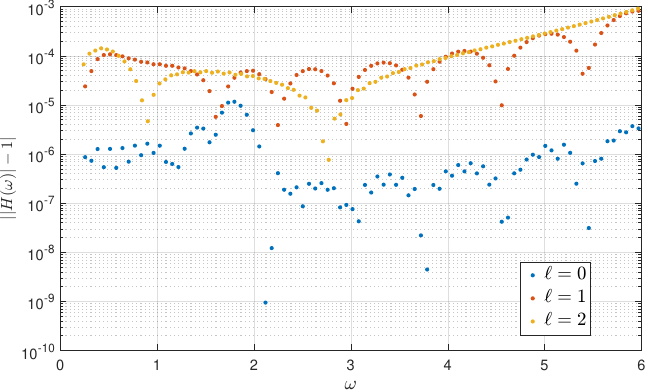}
\caption{Deviation $||H(\omega)|-1|$ from unitarity for $\ell=0,1,2$
  PT potential scattering ($A_0=1$, $\sigma_w=0.5,\,\omega_0=0$, $V_0=3$, $\sigma=3$, PDE $N=16384$).
  The logarithmic scale reveals residual deviations below
  $10^{-3}$.}
\label{fig:phaseshift_dev}
\end{figure}

\begin{figure*}[htbp] 
\centering
\includegraphics[width=0.95\textwidth]{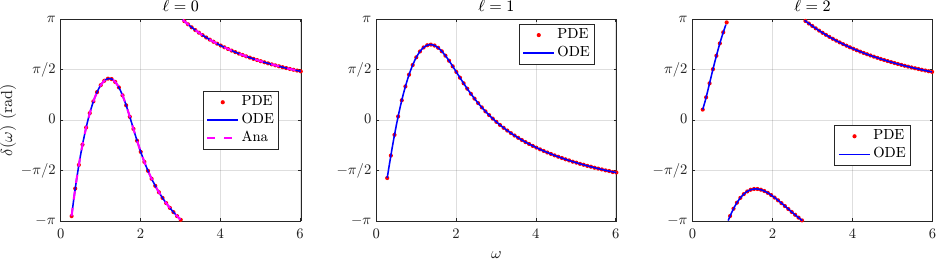}
\caption{Scattering phase shift $\delta(\omega)=\arg H(\omega)$ of PT potential scattering for
  $\ell=0,1,2$  ($A_0=1$, $\sigma_w=0.5,\,\omega_0=0$, $V_0=3$, $\sigma=3$).
  Here, we eliminate the modulo $2\pi$ ambiguity in the phase shift by choosing the branch $\delta(\omega)\in[-\pi,\pi)$.
  Red dots: PDE evolution ($N=8192$); blue solid: Chebyshev spectral method; magenta dashed ($\ell=0$ only): analytic
  formula~(\ref{eq:Sanalytic}).
  The agreement between all methods is ${\lesssim}10^{-3}$\,rad.}
\label{fig:phaseshift_phi}
\end{figure*}

The deviation $|H_\ell(\omega)|-1$ from unitarity, shown in Fig.~\ref{fig:phaseshift_dev} for $\ell=0,1,2$, remains at the $10^{-3}$ level across the resolved frequency band, confirming that the scattering is elastic to sub-percent accuracy.
The phase shift $\delta_\ell(\omega)=\arg H_\ell(\omega)$ extracted from the PDE evolution agrees with the analytic hypergeometric formula~(\ref{eq:Sanalytic}) for $\ell=0$ and with the reference Chebyshev spectral method for $\ell=1,2$ to within ${\sim}10^{-3}\,\mathrm{rad}$, as displayed in Fig.~\ref{fig:phaseshift_phi}.
These results demonstrate that the characteristic PDE method accurately resolves the full scattering information encoded in the complex transfer function $H_\ell(\omega)$, thereby validating the double-null framework as a reliable tool for quantitative scattering computations.

\subsection{Semi-linear $|\phi|^{n-1}\phi$}
\label{sec:semilinear_results}

\begin{figure*}[htbp] 
\centering
\includegraphics[width=\textwidth]{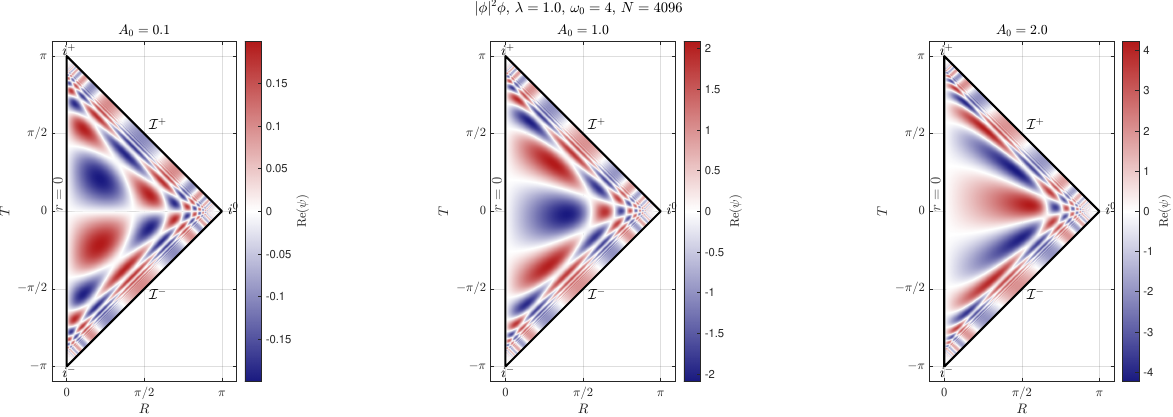}
\caption{Penrose diagrams of Re($\psi$) for the semi-linear $|\phi|^2\phi$
  theory with $\lambda=1$, $\omega_0=4$, $\sigma_w=5$, $N=4096$, and amplitudes
  $A_0=0.1,1,2$ (left to right).}
\label{fig:phi3_penrose}
\end{figure*}

The evolution for conformal scattering of the semi-linear case is globally second-order convergent including $\mathcal{I}^+$, and the Richardson extrapolation with $p=2$ further elevates the self-convergence rate of the $L_2$ norm to $Q_2{\gtrsim}4$, exhibiting a superconvergence (see Table~\ref{tab:semilinear_conv}).
Results of the cubic nonlinearity $|\phi|^{2}\phi$ are shown in Fig.~\ref{fig:phi3_penrose}.

\begin{table*}[htbp] 
\centering
\caption{Self-convergence rates of raw $\mathcal{R}_+$ and Richardson-extrapolated $\mathcal{R}^{[p]}_{+,N}$ for the semi-linear $|\phi|^{n-1}\phi$ theory ($\ell=0$, $\lambda=1$, $A_0=1$,
  $\sigma_w=5$, $\omega_0=4$).
  }
\label{tab:semilinear_conv}
\begin{tabular}{lcccccccccccc}
\toprule
 & \multicolumn{6}{c}{$Q_{2/\infty}(\mathcal{R}_+;N=2048)$} & \multicolumn{6}{c}{$Q_{2/\infty}(\mathcal{R}^{[2]}_{+,N};N=1024)$} \\
\cmidrule(lr){2-7}\cmidrule(lr){8-13}
$n$ & $Q_2$ & $\|e_{N,2N}\|_2$ & $\|e_{2N,4N}\|_2$ & $Q_\infty$ & $\|e_{N,2N}\|_\infty$ & $\|e_{2N,4N}\|_\infty$
       & $Q_2$ & $\|e_{N,2N}\|_2$ & $\|e_{2N,4N}\|_2$ & $Q_\infty$ & $\|e_{N,2N}\|_\infty$ & $\|e_{2N,4N}\|_\infty$ \\
\midrule
3 & \textbf{1.99} & $3.8\!\times\!10^{-6}$ & $9.5\!\times\!10^{-7}$ & \textbf{2.00} & $5.8\!\times\!10^{-3}$ & $1.5\!\times\!10^{-3}$
  & \textbf{5.50} & $2.1\!\times\!10^{-7}$ & $4.7\!\times\!10^{-9}$ & \textbf{3.97} & $9.2\!\times\!10^{-4}$ & $5.9\!\times\!10^{-5}$ \\
5 & \textbf{1.99} & $4.5\!\times\!10^{-6}$ & $1.1\!\times\!10^{-6}$ & \textbf{2.00} & $8.8\!\times\!10^{-3}$ & $2.2\!\times\!10^{-3}$
  & \textbf{4.43} & $3.9\!\times\!10^{-7}$ & $1.8\!\times\!10^{-8}$ & \textbf{3.97} & $1.8\!\times\!10^{-3}$ & $1.2\!\times\!10^{-4}$ \\
7 & \textbf{1.79} & $5.6\!\times\!10^{-6}$ & $1.0\!\times\!10^{-6}$ & \textbf{2.00} & $1.0\!\times\!10^{-2}$ & $2.5\!\times\!10^{-3}$
  & \textbf{4.60} & $6.5\!\times\!10^{-7}$ & $2.7\!\times\!10^{-8}$ & \textbf{4.06} & $3.1\!\times\!10^{-3}$ & $1.9\!\times\!10^{-4}$ \\
\bottomrule
\end{tabular}
\end{table*}

As a physical diagnostic, we check the deviation from energy conservation by computing
\begin{align}
  E_{\text{in}} &= \int_{-\infty}^{\infty} |\partial_v\mathcal{R}_-|^2\,\mathrm{d}v,
  &
  E_{\text{out}} &= \int_{-\infty}^{\infty} |\partial_u\mathcal{R}_+|^2\,\mathrm{d}u.
\end{align}
For the modulated Gaussian initial data \eqref{eq:init}, the incoming energy is evaluated analytically:
\begin{equation}
  E_{\text{in}} = A_0^2\sqrt{\pi}\,\sigma_w
    \left(\omega_0^2 + \frac{1}{2\sigma_w^2}\right).
  \label{eq:ein_analytical}
\end{equation}
The outgoing energy $E_{\text{out}}$ is computed from the numerical $\mathcal{R}_+(U)$ using the same FFT pre-processing pipeline described in Sec.~\ref{sec:pt_results} (truncation to the well-resolved region and spline resampling onto a uniform $u$-grid), followed by central-difference differentiation and trapezoidal integration.
Applied to the free-wave ($\lambda=0$) case---where the analytic incoming and outgoing fluxes are equal---this post-processing pipeline introduces an intrinsic energy error of ${\sim}3.43\times10^{-5}$ at $N=8192$, which we take as the baseline.
Table~\ref{tab:energy_dev} reports the energy deviation for each $(p,A_0)$ combination; all tested cases lie below $2.3\times10^{-4}$, well above the baseline but still at the $10^{-4}$ level.

\begin{table}[htbp] 
\centering
\caption{Energy conservation deviation $|E_{\text{out}} - E_{\text{in}}|/E_{\text{in}}$
  for the semi-linear $|\phi|^{n-1}\phi$ theory ($\lambda=1$, $\omega_0=4$,
  $N=8192$).}
\label{tab:energy_dev}
\begin{tabular}{cccc}
\toprule
$p$ & $A_0=0.1$ & $A_0=1$ & $A_0=2$ \\
\midrule
3 & $3.43\times10^{-5}$ & $3.57\times10^{-5}$ & $3.38\times10^{-5}$ \\
5 & $3.43\times10^{-5}$ & $1.35\times10^{-5}$ & $7.55\times10^{-5}$ \\
7 & $3.43\times10^{-5}$ & $1.69\times10^{-5}$ & $2.26\times10^{-4}$ \\
\bottomrule
\end{tabular}
\end{table}

\begin{figure*}[htbp] 
\centering
\includegraphics[width=\textwidth]{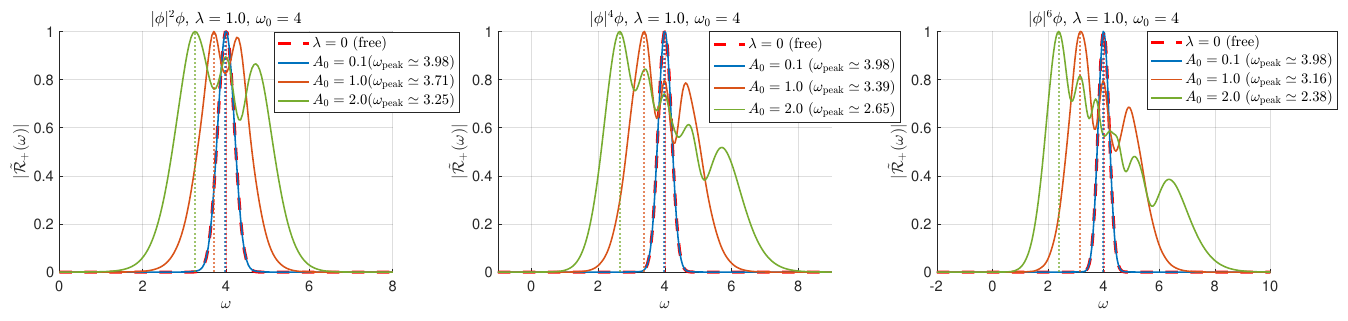}
\caption{Frequency spectra $|\tilde{\mathcal{R}}_+(\omega)|$ at $\mathcal{I}^+$
  for $|\phi|^{n-1}\phi$ theories with $n=3,5,7$ (left to right),
  $\lambda=1.0$, $\omega_0=4$, $N=4096$, and amplitudes $A_0=0.1,1,2$ and $\sigma_w=5$.
  Red dashed: the $\lambda=0$ (free wave) case is also displayed for comparison.
  These results show the spectral broadening due to the nonlinear coupling, known as the Kerr effect in nonlinear optics.}
\label{fig:phi3_spectrum}
\end{figure*}

Due to the nonlinear coupling $|\phi|^{n-1}\phi$, the modulus of the transfer function $H(\omega)$ depends on the coupling strength or amplitude, and may deviate from unity.
As shown in Fig.~\ref{fig:phi3_spectrum}, the $\lambda>0$ ($|\phi|^2\phi$) nonlinearity acts as a self-defocusing (negative Kerr) medium~\cite{Boyd2020}: the spectral peak at the carrier frequency $\omega_0$ is suppressed relative to the free wave, while the pulse develops a multi-peak structure characteristic of self-phase modulation.
The strongest spectral peak away from $\omega_0$ is consistently redshifted ($\omega<\omega_0$), reflecting the dominant contribution of the pulse leading edge to the nonlinear phase accumulation.
As the coupling strength or amplitude increases ($n=5,7$ or larger $A_0$), the spectral broadening becomes more pronounced, generating additional sidebands whose amplitude grows with the nonlinear phase accumulation.

\section{Conclusion and outlook}
\label{sec:conclusion}

We have presented a numerical framework for characteristic conformal scattering of massless scalar waves in Minkowski spacetime, formulated as a Goursat problem of the physical wave equation in compactified double-null coordinates $(U,V)$.
The framework has been applied to three physical scenarios: free wave propagation, linear scattering off the PT potential, and the semi-linear $|\phi|^{n-1}\phi$ wave equation ($n=3,5,7$).
The free-wave case, though the simplest, already exhibits the essential stability and convergence properties of our numerical schemes.
For $\ell=0,1$, the explicit stencil with center value $\psi_{\rm c}\simeq(\psi_{\text{E}}+\psi_{\text{W}})/2$ achieves global $O(\Delta^2)$ convergence.
For $\ell\ge2$, the implicit stencil with $\psi_{\rm c}\simeq(\psi_{\text{N}}+\psi_{\text{S}})/2$ is unconditionally stable under local CFL analysis.
The convergence of the implicit stencil is reduced to $O(\Delta)$ on $\mathcal{I}^+$.
Nevertheless, it can be partially remedied by Richardson extrapolation (the convergence rate $Q_2$ improved to ${\sim}1.5$).
For PT scattering, the scattering phase shift $\delta_\ell(\omega)$ is extracted and cross-validated against an analytic $S$-matrix ($\ell=0$) and a Chebyshev spectral collocation method working in the frequency domain ($\ell=1,2$), with agreement at the ${\sim}10^{-3}\,\mathrm{rad}$ level.
The elasticity of the linear scattering is confirmed to sub-percent accuracy.
The semi-linear case demonstrates that the explicit stencil preserves global $O(\Delta^2)$ accuracy on $\mathcal{I}^+$ under spherical symmetry and successfully captures the physical signatures of a self-defocusing Kerr nonlinearity, including self-phase modulation and spectral broadening.

The main limitation of the present work is the poor convergence on future null infinity $\mathcal{I}^+$ when $\ell\ge2$, which is essential for the gravitational scattering. 
This requires a dedicated treatment of spatial infinity $i^0$, following approaches such as \cite{Frauendiener:2013vda,FrauendienerHennig:2016,FrauendienerHennig:2017,Macedo:2018txl,Hennig:2020rns}.
The methods developed here provide the foundation for the conformal scattering problem in black hole spacetimes, to be addressed in a forthcoming paper~\cite{He2026}.
Our long-term goal is to achieve fully nonlinear numerical global evolution of conformal scattering for gravitational waves in black hole spacetimes.

\begin{acknowledgments}
We thank Xiao-Ning Wu for introducing the development of conformal scattering in mathematical relativity.
This work is partly supported by the National Key Research and Development Program of China (Grant No. 2021YFC2203001).
This work is supported in part by the National Natural Science Foundation of China under Grants No. 12035016, No. 12075026, No. 12275350, No. 12375048, No. 12375058, No. 12361141825, No. 12447182 and No. 12575047.

The authors acknowledge the use of AI-assisted coding and writing tools during the preparation of this manuscript: DeepSeek V4 (Pro and Flash) for primary code development and manuscript editing; MiMo V2.5-Pro for literature survey and preliminary draft of the introduction; and GLM 5.1 and 5.2 for assistance in the design of the numerical scheme.
All AI outputs were reviewed, validated, and directed by the authors, who bear full responsibility for the scientific content.
\end{acknowledgments}

\section*{Data Availability}
The data that support the findings of this article are openly available \cite{He_-_DN_CScattering_M_2026}.
\appendix
\section{Numerical computation of the phase shift in frequency domain}
\label{app:cheb}
For $\ell\ge1$, the centrifugal term $\ell(\ell+1)/r^2$ is present but can be eliminated by the Frobenius substitution $\psi_\omega=r^{\ell+1}\hat\psi$, which yields a regular ODE with no $1/r^2$ singularity:
\begin{equation}
  r\hat\psi''+2(\ell+1)\hat\psi'+r\bigl[\omega^2-V_0\operatorname{sech}^2
  (r/\sigma)\bigr]\hat\psi=0.
  \label{eq:frobenius_ode_app}
\end{equation}
Equation~\eqref{eq:frobenius_ode_app} is solved on $[0,r_0]$ with the regularity condition $\hat\psi'(0)=0$ and normalization $\hat\psi(0)=1$.
The exterior region $r\in[r_0,\infty)$ is mapped to $z=1/r\in[0,z_0]$ with $z_0=1/r_0$, and the outgoing asymptotic behavior $\psi_\omega\sim \mathrm{e}^{\mathrm{i}\omega r}$ is factored out via $\psi_\omega=\mathrm{e}^{\mathrm{i}\omega r}\tilde\psi_{\rm out}(z)$, giving
\begin{widetext}
\begin{equation} 
  z^2\tilde\psi_{\rm out}''+2(z-\mathrm{i}\omega)\tilde\psi_{\rm out}'+
  \left[-\ell(\ell+1)-
  \frac{V_0}{z^2}
  \operatorname{sech}^2\left(\frac{1}{\sigma z}\right)\right]
  \tilde\psi_{\rm out}=0,
  \label{eq:exterior_ode_app}
\end{equation}
\end{widetext}
with Robin boundary condition $-2\mathrm{i}\omega\tilde\psi_{\rm out}'(0) -\ell(\ell+1)\tilde\psi_{\rm out}(0)=0$ and normalization $\tilde\psi_{\rm out}(0)=1$.
For real frequencies, the incoming solution is obtained by complex conjugation: $\tilde\psi_{\rm in}(z)=\overline{\tilde\psi_{\rm out}(z)}$.

Both ODEs are discretized on Chebyshev collocation grids of $N+1$ points via the differentiation matrix $D$~\cite{Trefethen00}.  The degenerate boundary rows are replaced by their respective boundary conditions; the normalization conditions replace one interior row each, yielding two well-posed linear systems.  
$C^1$ matching at $r=r_0$ gives
\begin{equation}
  r_0^{\ell+1}\hat\psi(r_0)=A_{\rm out}\tilde\psi_{\rm out}(z_0)\mathrm{e}^{\mathrm{i}\omega r_0}
  +A_{\rm in}\tilde\psi_{\rm in}(z_0)\mathrm{e}^{-\mathrm{i}\omega r_0},
\end{equation}
and the same for the first derivative, from which the $S$-matrix element $S_\ell(\omega)=A_{\rm out}/A_{\rm in}$ and transfer function $H_\ell(\omega)=-\overline{S_\ell(\omega)}$ are obtained.
Since $\hat\psi$ and $\tilde{\psi}_{\rm out}$ are smooth functions, the Chebyshev spectral method converges exponentially with $N$.
Results shown in Fig.~\ref{fig:phaseshift_phi} are obtained with $N=64$.

\newpage
\bibliography{refs}

\end{document}